\documentclass{PoS}

\title{Volume reduction through perturbative Wilson loops}

\ShortTitle{Volume reduction through perturbative Wilson loops}

\author{\speaker{Margarita Garc\'{\i}a P\'erez}
      \\
      Instituto de F\'{\i}sica Te\'orica UAM/CSIC, \\ Universidad Aut\'onoma de Madrid, Nicol\'as Cabrera 13-15, Cantoblanco, 28049 Madrid, Spain\\
      E-mail: \email{margarita.garcia@uam.es}}

\author{Antonio Gonz\'alez-Arroyo\\
        Instituto de F\'{\i}sica Te\'orica UAM/CSIC and Departamento de F\'{\i}sica Te\'orica C-15,  \\Universidad Aut\'onoma de Madrid, Cantoblanco, 28049 Madrid, Spain\\
        E-mail: \email{antonio.gonzalez-arroyo@uam.es}}

\author{Masanori Okawa\\
Graduate School of Science and Core of Research for the Energetic Universe, \\
Hiroshima University, Higashi-Hiroshima, Hiroshima 739-8526, Japan, \\
E-mail: \email{okawa@sci.hiroshima-u.ac.jp}}

\abstract{
We derive the perturbative expansion of Wilson loops to order $g^4$ in a $SU(N)$ lattice gauge theory with twisted boundary conditions. 
Our expressions show that the thermodynamic limit is attained at infinite N for any number of lattice sites and allow to 
quantify the deviations from volume independence at finite large N as a function of the twist. }

\FullConference{34th annual International Symposium on Lattice Field Theory\\
		 24-30 July 2016\\
		 University of Southampton, UK}

\newcommand{\HG}{\hat{\Gamma}}
\newcommand{\be}{\begin{equation}}
\newcommand{\ee}{\end{equation}}
\newcommand{\ba}{\begin{array}}
\newcommand{\ea}{\end{array}}
\newcommand{\baa}{\begin{array}}
\newcommand{\eaa}{\end{array}}
\newcommand{\bea}{\begin{eqnarray}}
\newcommand{\eea}{\end{eqnarray}}
\newcommand{\half}{\frac{1}{2}}

\newcommand{\Tr}{\mathrm{Tr}}

\newcommand{\kb}{\bar{k}} 

\newcommand{\hL}{\hat{L}}

\newcommand{\LL}{L}
\newcommand{\kbar}{\bar k}

\newcommand{\tw}{\tilde{W}}
\newcommand{\RR}{\mathcal{G}}
\newcommand{\Le}{L_{\mathrm{eff}}}
\newcommand{\Ve}{V_{\mathrm{eff}}}
\newcommand{\pbc}{\mathrm{PBC}}
\newcommand{\tbc}{\mathrm{TBC}}

\begin{document}

\section{Introduction}

Twisted boundary conditions (TBC) \cite{'tHooft:1979uj,'tHooft:1980dx} have been  a useful tool for performing perturbative calculations
at finite volume~\cite{GonzalezArroyo:1982hz}-\cite{Snippe:1997ru}. Their main advantage over periodic boundary conditions 
is the discreteness of the zero-action solution, avoiding the complications associated to the existence of the toron 
valley~\cite{GonzalezArroyo:1981vw}-\cite{vanBaal:1986ag}. The twist is also essential to the formulation of 
the TEK model~\cite{GonzalezArroyo:1982ub,GonzalezArroyo:1982hz}, a successful proposal for implementing the idea of volume reduction put forward by 
Eguchi and Kawai~\cite{Eguchi:1982nm}. In this work, we will focus on computing 
the expectation value of Wilson loop operators in lattice perturbation theory for the case of the symmetric twist 
introduced in~\cite{GonzalezArroyo:2010ss}. 
Our results will be compared with the ones obtained at infinite volume~\cite{DiGiacomo:1981wt}-\cite{Alles:1993dn}
and with those derived at finite volume with periodic boundary conditions (PBC)~\cite{Heller:1984hx}.
The latter is obtained by expanding around the trivial vacuum and neglecting the contribution  of the 
zero-momentum modes. The correct PBC calculation can be done along the lines explained in Ref.~\cite{Coste:1985mn}. 
Our TBC calculation will allow to test volume reduction at large N and to quantify the size of finite N corrections.

\section{Methodology}
We will consider a $SU(N)$ gauge theory on a lattice of size $L^4$ with twisted boundary conditions, defined by the Wilson action 
\be
S = b N \sum_n \sum_{\mu \nu } [N - Z_{\mu \nu }(n) \Tr (U_\mu(n)
U_\nu(n+\hat \mu ) 
U_\mu^\dagger(n+\hat \nu ) U_\nu^\dagger(n)) ].
\ee
In this expression the link variables are periodic and the twist is introduced through the factors
$Z_{\mu \nu}(n)=Z^*_{\nu \mu}(n)$ belonging to the center of the gauge group. 
We will impose the so-called symmetric twist. It requires $N$ to be the square of an integer, $N= \hL^2$, and is implemented on the lattice by
taking the twist-carrying factors equal to one at all plaquettes except for one at each $\mu\nu$ plane where
$Z_{\mu \nu}= \exp (2 \pi i \epsilon_{\mu \nu} k/\hL)$, with $\epsilon_{\mu \nu}=\theta(\nu-\mu) - \theta(\mu-\nu)$, and $k$  an integer 
coprime with $\hL$.   

We will focus on computing in perturbation theory the expectation values of $R\times T$ Wilson loops, given on the twisted lattice by:
\be
\label{expectWL}
W_{R,T}(b,N,\LL,k)=\frac{1}{N} Z(R,T)\langle \Tr(U(R,T)) \rangle
\ee
where $Z(R,T)$ is the product of all the $Z_{\mu \nu}(n)$ factors from plaquettes contained within the loop.

\subsection{The perturbative expansion}

The first step is to expand the link variables around the zero-action solutions of the twisted Wilson action: 
\be
\label{expansion}
U_\mu(n)= e^{-i g A_\mu(n)}  \Gamma_\mu(n),
\ee
With our choice of twist, all the $\Gamma_\mu(n)$ matrices can be chosen equal to the identity except for the ones at the edges of the lattice that
are given by:  $\Gamma_\mu(n_\mu=\LL-1)\equiv\Gamma_\mu$, with
$\Gamma_\mu$ a solution of the equation $\Gamma_\mu \Gamma_\nu=Z_{\nu \mu} \Gamma_\nu \Gamma_\mu$. 
The gauge potential $A_\mu(n)$ satisfies the boundary conditions $A_\mu(n+L \hat \nu) = \Gamma_\nu A_\mu(n) \Gamma_\nu^\dagger$, 
which can be easily implemented in momentum space by taking \cite{GonzalezArroyo:1982hz}:
\be
\label{fourier}
A_\mu(n)=  \frac{1}{L^2} \sum'_{q} e^{i
q (n+\half)} \hat{A}_\mu(q)  \HG(q). 
\ee
In this expansion momentum is quantized as $q_\mu  = 2 \pi m_\mu /\Le$, with $m_\mu$ an integer defined modulo $\Le \equiv \LL \hL$,
excluding values for which 
$m_\mu = 0 \, (\bmod \hL)$, for all $\mu$. The precise definition of the momentum dependent Lie algebra basis is not relevant here, it suffices to know 
that the $\HG(q)$ matrices  satisfy the commutation relations:
\be
\label{eq.comm}
[\HG(p), \HG(q)] = i \, F(p, q , -p-q) \, \HG(p+q) , \qquad F(p,q,-p-q)= -\sqrt{\frac{2}{N}}  \,  
\sin\left(\frac{ \theta_{\mu \nu}}{2} \, p_\mu q_\nu
\right) \, ,
\ee
where we have introduced the twist-dependent antisymmetric tensor:
\be
\label{theta}
\theta_{\mu \nu} =    \frac{ \LL^2 \hL^2 } {2\pi} \times  \, \tilde
\epsilon_{\mu \nu} \, \frac{\bar k}{  \hL}  \, ,
\ee
with the integer $\kb$ defined by the condition: $\kb k = 1 \, (\bmod  \hL )$, and $\tilde
\epsilon_{\mu \nu} \epsilon_{\nu\rho} = \delta_{\mu \rho}$.

 It is now easy to derive the Feynman rules for the twisted Wilson action. One just has to follow the standard procedure in lattice
perturbation theory. Let us  mention for example that in Feynman gauge, chosen
for our calculation, the propagators of the gauge and ghost fields read:
\be 
\label{eq:prop}
P_{\mu \nu} (p, q) = 
\delta_{\mu \nu} \,  \delta(q+p)  \, {1 \over \widehat q^2} \, , \qquad P_{GH} (p, q) =
 \delta(q+p)  \, {1 \over \widehat q^2} \, ,
\ee
with $\widehat q_\mu = 2 \sin(q_\mu/2)$.  
Notice that both propagators are identical to those on a finite lattice with effective size $\Le$. This is the first manifestation of
volume reduction in perturbation theory. 

The perturbative expansion of the Wilson loop is obtained by inserting the expansion of the link variables  
into  Eq. (\ref{expectWL}), using the 
the Baker-Campbell-Haussdorf formula to rewrite it as:
\be
 Z(R,T) U(R,T) =  \exp \Big \{-i g \Big (\RR^{(1)} + g^2 \RR^{(2)} + g^3 \RR^{(3)} + {\cal O}(g^4)\Big )\Big \},
\ee
The trace of the loop is then given by:
\bea
{1 \over N} \Tr (Z(R,T) U(R,T)) &=& 
 1 -{g^2 \over 2 N} \Tr (\RR^{(1)})^2 - {g^3 \over N} \Big ( \Tr (\RR^{(1)} 
\RR^{(2)}) - {i \over 3!} \Tr (\RR^{(1)})^3 \Big ). 
 \\
&-& {g^4 \over N} \Big ( \half \Tr (\RR^{(2)})^2
 + \Tr (\RR^{(1)} \RR^{(3)}) - {1 \over 4 !} \Tr (\RR^{(1)})^4
\Big)
\nonumber
\eea
It is important to remark that the imaginary part of the loop does not vanish since the twist breaks 
CP invariance, it nevertheless tends to zero in the infinite volume limit. 

We will not give here explicit expressions for the different terms, details will soon appear in \cite{GPGAO}. They are as usual derived 
by inserting the momentum expansion of the gauge potential and the expressions for its $n$-point Green functions. 
Instead, we will focus on discussing the results and analyzing the volume and $N$ dependence of the different contributions.

\section{Results}
In this section, we will present our results for the perturbative expansion of the logarithm of the Wilson
loop in terms of the 't Hooft coupling, $\lambda = 1/b \equiv 2 N^2 / \beta$, up to order $\lambda^2$:
\be
\log\left(W_{R,T}(b,N,L,k)\right)=-\lambda
\tilde{W}^{(R\times T)}_1(N,L,k)- \lambda^2
\tilde{W}^{(R\times T)}_2(N,L,k)+ \ldots
\ee
We will discuss first the leading order result and proceed next to present the results at order $\lambda^2$.

\subsection{The Wilson loop at ${\cal O} (\lambda)$}

The expression for the Wilson loop at leading order in $\lambda$ is given by:
\be
\tilde{W}^{(R\times T )}_1 (N,L,k)=    
{1 \over   \Le^4 } \sum_q^\prime  \frac{\sin(R q_\mu /2 ) \sin(T q_\nu /2 )}{\widehat q_\mu \widehat q_\nu}  \  \  \frac{\widehat q_\mu^2 + \widehat q_\nu^2}{ \widehat q^2} \equiv {1 \over   \Le^4 } \sum_q^\prime P_{\mu \nu}^{(R\times T )}\, .
\ee
This formula can be related to the one for PBC given  in Ref. ~\cite{Heller:1984hx} by the substitution:
\be
{1 \over   \Le^4 } \sum'_q \longrightarrow   \frac{N^2-1} {N^2 L^4} \sum_{q\ne 0}
\ee 
Up to the $N$ dependent factors, the main difference between both expressions resides in the set of momenta included in the sum. While momenta in the 
twisted lattice are quantized in units of $\Le$, those in the periodic box go in units of $L$. We will denote these two sets of lattice momenta 
by $\Lambda_{\Le}$ and $\Lambda_{L}$ respectively. 
Writing $\Lambda_{L}^\prime = \Lambda_L - \{0\}$,  the periodic Wilson loop is given at this order by:
\be
\tw_1^\pbc (N,L,k=0)= \Big (1 - \frac{1}{N^2}\Big )  F_1(L) ; \qquad F_1(L) =  \frac{1}{L^4} \sum_{q \in \Lambda_L^\prime} P_{\mu \nu}^{(R\times T )}
\ee 
The function $F_1(L)$ vanishes on the one-point lattice $F_1(1) = 0$, and behaves at large $L$ as:
\be
F_1(L) = F_1(\infty) - \frac{R^2 T^2}{8 L^4} + {\cal O} (1/L^6).
\ee
For TBC, the set of allowed momenta belongs instead to the lattice $\Lambda_{\Le} \backslash \Lambda_{L}$ and we obtain:
\be
\label{eq.w1tbc}
\tw^\tbc_1(N, \LL, k\ne 0)= F_1(\Le)-  \frac{1}{N^2} F_1(\LL).
\ee
It is interesting to analyze both expressions in the two limiting cases: $L$ or $N$ going to infinity.
In the thermodynamic limit they agree, leading to the correct infinite volume result:
\be
\tw_1(N, L=\infty)= \Big (1-  \frac{1}{N^2}\Big ) F_1(\infty).
\ee
The two expressions differ instead in the large $N$ limit. For TBC, this limit leads to $F_1(\infty)$, recovering the large $N$, infinite 
volume result regardless of the size, $\LL$, of the lattice. This is however not the case for PBC,
where the large $N$  expression tends to $F_1(L)$, retaining finite volume effects.

\begin{figure}[t]
\begin{center}
\includegraphics[width=.67\linewidth]{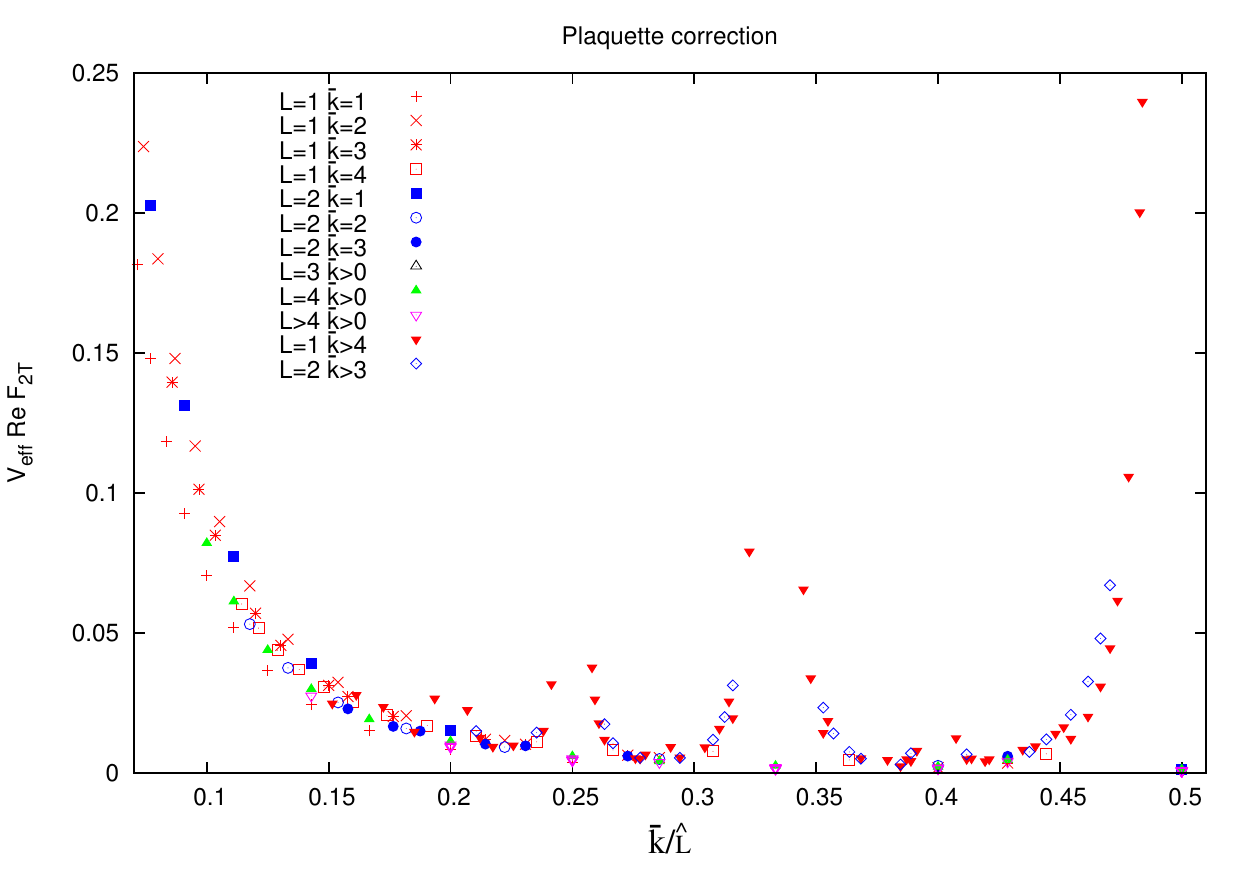}
\caption{The real part of $\Ve F_{2T}(\Le)$ for the plaquette in the 12 plane is plotted as a function of
$\kbar/\hL$.
}
\label{figf2tA}
\end{center}
\end{figure}

\subsection{The Wilson loop at ${\cal O} (\lambda^2)$}

In the previous subsection we have seen that, at leading order in $\lambda$, the large $N$ twisted Wilson loop tends to the 
infinite volume result irrespective of the lattice size. This is a clear manifestation of volume reduction. The corrections to this behaviour
are, according to Eq. (\ref{eq.w1tbc}), suppressed as $1/N^2$. We will now see that this is also the case for the ${\cal O} (\lambda^2)$ term.

One can show that, at second order in the 't Hooft coupling, the periodic loop is given by:
\be
\tw^\pbc_2(N, \LL, k=0)= \Big (1-\frac{1}{N^2}\Big ) F_2(\LL)+ \Big (1-\frac{1}{N^2}\Big )^2 F_W(\LL),
\ee
expressed in terms of two new functions that can be related to the ones appearing in ~\cite{Heller:1984hx} through:
\be
\bar{W}_2(\LL)=F_1(\LL), \qquad Y(\LL)=2 F_W(\LL)-F_1^2(\LL), \qquad X(\LL)= F_2(\LL)+ \frac{1}{6} Y(\LL).
\ee
It is also easy to see that $F_W$ is related to the function $F_1$ appearing at lowest order by:
\be
F_W(L)= \frac{1}{8}\Big (1-\frac{1}{L^4}\Big )F_1(L).
\ee
For TBC instead, the different contributions lead to:
\be
\label{MAIN}
\tw^\tbc_2(L,N, k)= \Big (F_2(\Le)-\frac{1}{N^2}F_2(\LL)\Big ) + 
\frac{1}{8}\Big (1-\frac{1}{N^2}\Big )\Big (F_1(\Le)-\frac{F_1(L)}{N^2}\Big ) 
+ F_{2T}(\LL,N,k). 
\ee
The last term involves a complex function specific of the twisted case. We don't have enough space to describe how this formula
is derived, a detailed description will be given in \cite{GPGAO}. Let us just mention that the key point in this decomposition is to
write the terms depending quadratically on the structure constants as: $NF^2= 1- \cos(\theta_{\mu \nu} p_\mu q_\nu) $. The function $F_{2T}$ 
involves the terms proportional to the cosine and it is the only one 
exhibiting an explicit dependence on the flux $k$.

We can now repeat the analysis done for the leading order term and look at the large $L$ or large $N$ limits of these expressions. 
For large $L$:
\be 
\tw^\tbc_2(\infty,N, k) = \Big (1-\frac{1}{N^2}\Big ) F_2(\infty)+ \frac{1}{8} \Big (1-\frac{1}{N^2}\Big )^2 F_1(\infty ) + F_{2T}(\infty,N,k). 
\ee
This recovers the infinite volume result, provided $F_{2T}$ goes to zero in the thermodynamic limit, a property that can be explicitly derived.
Taking instead the large $N$ limit we obtain:
\be 
\tw^\tbc_2(\LL,\infty, k) = F_2(\infty)+ \frac{1}{8} F_1(\infty ) + F_{2T}(\LL,\infty,k).
\ee
If $F_{2T}(\LL,N,k)$ goes to zero for large $N$, we recover volume reduction: the large $N$ limit of the twisted theory reproduces the 
large $N$ thermodynamic result irrespective of the lattice size.

We have computed the $F_{2T}$ function for several values of $N$, $\LL$ and $\kbar$. 
In Fig. \ref{figf2tA} we show an 
example of the results obtained for the plaquette in the 12 plane~\footnote{                                    
In analogy to the case of CP, the twist induces a breakdown of rotational invariance.                                               
For our choice of twist this leaves a residual symmetry grouping planes into two different sets 
that have a similar, although not identical,  behaviour.}.  We display as a function of  $\kbar/\hL$ the real part of $F_{2T}$ multiplied by the effective volume 
$\Ve= \LL^4 N^2$. 
The results confirm that this function vanishes when either $N$ or $L$ goes to infinity. 
The same conclusion is obtained for the imaginary part. 

Our numerical code allows also for a very precise determination of the functions $F_1$ and $F_2$.
In the case of $F_2$, the dependence on $\LL$ is well described by a formula:
\be
\label{f2formula}
F_2(\LL)=F_2(\infty)- \frac{R^2
T^2(\gamma_2+\gamma_2'\log(\LL))}{\LL^4}+ \ldots
\ee
We have computed $F_2$ up to a value of $\LL=34$, allowing to determine the infinite volume coefficient with high precision. Our results, 
presented in table \ref{tableI}, are
consistent with the  precise  results of Ref.~\cite{Alles:1993dn} for the plaquette and improve
significantly the results published for larger loops. 

\begin{table}
\begin{center}
\begin{tabular}{||l||c|c|c|c||}\hline \hline
LOOP & $1 \times 1$ & $2 \times 2$ & $3 \times 3$ & $4 \times 4$   \\ \hline \hline
$F_1(\infty)$ &  0.125 & 0.34232788379 & 0.57629826424  &  0.81537096352  \\ \hline
$F_2(\infty)$ & -0.0027055703(3) & -0.00101077(1) & 0.00295130(2) & 0.0076217(1)
 \\ \hline 
\end{tabular}
\caption{ Values at infinite volume of the functions $F_1$ and $F_2$
defined in the text. 
  }
\label{tableI}
\end{center}
\end{table}

\section{Conclusions}

We have derived the pertubative expansion of the Wilson loop up to order $\lambda^2$ for the Wilson lattice action
and twisted boundary conditions. Our results depend on three functions, two of which appear also in the case of periodic 
boundary conditions studied in \cite{Heller:1984hx}. The third, $F_{2T}$, is specific of the twisted case and encodes all the dependence on
the flux $k$. We have shown that volume independence holds if this function goes to zero in the large $N$ limit. Our numerical values up to $\Ve = 34^4$, indicate
that this is indeed the case, with $F_{2T}$ vanishing as $1/(N^2 \LL^4)$.  Our formulas also allow to quantify the deviations from 
volume independence at finite $N$. For the one-site TEK lattice, for instance:
\be
\tw^\tbc_1(1,N, k)= \tw^\pbc_1(\sqrt{N},\infty, 0), \qquad \tw^\tbc_2(1,N, k)=  \tw^\pbc_2(\sqrt{N},\infty, 0) + F_{2T}(1,N,k),
\ee
showing that, at this order, finite volume effects in the TEK model are analogous to those of a periodic  
lattice of size $\sqrt{N}$, in the limit of infinite number of colours. This holds up to the corrections induced by $F_{2T}$ which come from 
non-planar diagrams and are suppressed as $1/N^2$.

\acknowledgments{
We acknowledge financial support from the
grants FPA2012-31686, FPA2012-31880,  FPA 2015-68541-P
and the Spanish MINECO's ``Centro
de Excelencia Severo Ochoa'' Programme under grant
SEV-2012-0249. M. O. is supported by the Japanese MEXT grant No
26400249 and the MEXT program for promoting the enhancement of research
universities. We acknowledge the use of the
HPC resources at IFT.  
}

\bibliographystyle{JHEP}

\providecommand{\href}[2]{#2}\begingroup\raggedright\endgroup

\end{document}